\documentclass{article}
\usepackage{natbib}
\setcitestyle{numbers,square,comma}

\usepackage[preprint]{neurips_2024}
\usepackage{etoolbox}
\newtoggle{isSubmission}
\togglefalse{isSubmission}

\providecommand{\keywords}[1]
{
  \textbf{\textit{Keywords: }} #1
}

\usepackage[utf8]{inputenc} 
\usepackage[T1]{fontenc}    
\usepackage{hyperref}       
\usepackage{url}            
\usepackage{booktabs}       
\usepackage{amsfonts}       
\usepackage{nicefrac}       
\usepackage{microtype}      
\usepackage[table]{xcolor}         
\usepackage{wrapfig}
\usepackage{authblk}

\title{Towards AI-\textbf{$45^{\circ}$} Law: A Roadmap to Trustworthy AGI}

\author{Chao Yang$^{1}$ \quad Chaochao Lu$^{1}$  \quad  Yingchun Wang$^{1}$ \quad Bowen Zhou$^{1,2}$\thanks{Corresponding author}\vspace{0.03in} \\
$^1$Center for Safe $\&$ Trustworthy AI, Shanghai AI Laboratory \quad $^2$Tsinghua University \\
}

\date{\today}

\usepackage{graphics}
\usepackage{graphicx}
\usepackage{subcaption}
\usepackage{caption}
\usepackage{amssymb}
\usepackage{algorithm}
\usepackage{algpseudocode}
\usepackage{wrapfig}
\usepackage{longtable}
\usepackage{fancyvrb}
\usepackage{empheq}
\definecolor{metacolor}{HTML}{0064E0}
\hypersetup{
  colorlinks,
  linkcolor=metacolor,
  citecolor=metacolor,
  urlcolor=metacolor
}

\usepackage{amsmath,amsfonts,bm}

\def\eqref#1{equation~\ref{#1}}

\def\1{\bm{1}}

\DeclareMathAlphabet{\mathsfit}{\encodingdefault}{\sfdefault}{m}{sl}
\SetMathAlphabet{\mathsfit}{bold}{\encodingdefault}{\sfdefault}{bx}{n}

\begin{document}
\maketitle

\begin{abstract}
Ensuring Artificial General Intelligence (AGI) reliably avoids harmful behaviors is a critical challenge, especially for systems with high autonomy or in safety-critical domains. Despite various safety assurance proposals and extreme risk warnings, comprehensive guidelines balancing AI safety and capability remain lacking.
In this position paper, we propose the \textit{AI-\textbf{$45^{\circ}$} Law} as a guiding principle for a balanced roadmap toward trustworthy AGI, and introduce the \textit{Causal Ladder of Trustworthy AGI} as a practical framework. This framework provides a systematic taxonomy and hierarchical structure for current AI capability and safety research, inspired by Judea Pearl's ``Ladder of Causation''.
The Causal Ladder comprises three core layers: the Approximate Alignment Layer, the Intervenable Layer, and the Reflectable Layer. These layers address the key challenges of safety and trustworthiness in AGI and contemporary AI systems.
Building upon this framework, we define five levels of trustworthy AGI: perception, reasoning, decision-making, autonomy, and collaboration trustworthiness. These levels represent distinct yet progressive aspects of trustworthy AGI. Finally, we present a series of potential governance measures to support the development of trustworthy AGI.\footnote{In this paper, trustworthiness is generally considered a broad form of safety, and no explicit distinction is made between the two. However, in some contexts, safety and trustworthiness are treated as distinct: safety involves assurance of correct behavior, while trustworthiness refers to user confidence in the system's decision-making. In such cases, different terms or both may be used depending on the context.}
\end{abstract}

\section{Crippled AI: The Imbalance Between AI Capability and Safety}

\subsection{Rapid Development of AI Capabilities}  
Artificial intelligence (AI) is experiencing a period of rapid advancement, driven by innovations in scaling laws \cite{rosenfeld2021scaling}, as well as breakthroughs in model architecture and computational resources \cite{kaplan2020scaling}. These developments have led to AI systems like ChatGPT \cite{brown2020language} and GPT-4 \cite{OpenAI2023-bl}, which demonstrate remarkable abilities in understanding and generating human-like language \cite{rayhan10natural}. These systems push the boundaries of AI's potential, approaching or surpassing human-level performance across various domains, including natural language processing \cite{rayhan10natural} and creative problem-solving \cite{renze2405self}. However, this rapid progress also presents significant risks \cite{statement}, as the development and deployment of these advanced systems often outpace the implementation of safety measures. The rapid growth in AI capabilities, coupled with the slow evolution of corresponding safety protocols \cite{mu2024rule}, highlights a critical imbalance that may hinder the responsible and reliable use of these technologies.

\subsection{Limitations of Current AI Safety Measures}

The current landscape of AI safety is predominantly characterized by a ``reactive approach'' approach, where safety measures are implemented only after models are developed and vulnerabilities are identified \cite{ayyamperumal2024current}. These measures are often domain-specific, tailored to particular applications, and lack the flexibility to be applied across diverse AI systems. For example, adversarial defense techniques \cite{lee2024deepfakes} designed to secure image recognition models are not easily adaptable to tasks such as speech recognition or natural language processing. Furthermore, the absence of a unified framework or standardized guidelines for AI safety exacerbates this issue, resulting in fragmented efforts that fail to comprehensively address the broad spectrum of potential risks. Various threats \cite{brundage2018malicious, motlagh2024large}, including adversarial attacks \cite{perez2022red, hong2024curiosity}, data poisoning \cite{carlini2021extracting}, and model theft \cite{zeng2023huref, zhang2024reef}, each require unique defensive strategies. Yet these strategies frequently operate in isolation, lacking the coordination and integration needed to tackle the increasingly complex and interconnected challenges posed by advanced AI systems \cite{huang2024pixels, huang2023flames, lawless2017evaluations, carlsmith2022power}.

\subsection{The Crippled State of AI Development}
The reactive and fragmented nature of AI safety practices has left the field in a state that can be described as ``crippled AI''. While tools like red teaming \cite{beuteldiverse, pavlova2024automated}, watermarking \cite{liu2024survey, watermarklearning}, and safety guardrails \cite{oh2024uniguard} provide some protection, their effectiveness is limited. Red teaming, for example, can uncover specific vulnerabilities through simulated attack scenarios, but it cannot address the full spectrum of potential threats. Similarly, watermarking techniques to identify AI-generated content can be tampered with or circumvented, rendering them unreliable at scale. Post-hoc safety measures such as guardrails \cite{gu2024mllmguard, liu2025mm, liu2023query, dong2024attacks, liu2024don, li2024salad} often reduce system flexibility and usability, while evaluation frameworks lack the theoretical grounding needed for comprehensive assessments. These shortcomings reflect a broader issue: the rapid development of AI capabilities is outpacing efforts to ensure their safety, resulting in powerful but fragile systems. This imbalance risks undermining public trust, limiting the practical adoption of AI, and creating systems that could inadvertently cause harm or operate in ways misaligned with human values. To bridge this gap, a proactive, unified, and scalable approach to AI safety is essential to ensure that the promise of AI is realized without compromising ethical and security considerations.
\section{AI-$45^{\circ}$ Law}
\begin{figure}[t]
      \centering
      \includegraphics[width=1.0\linewidth]{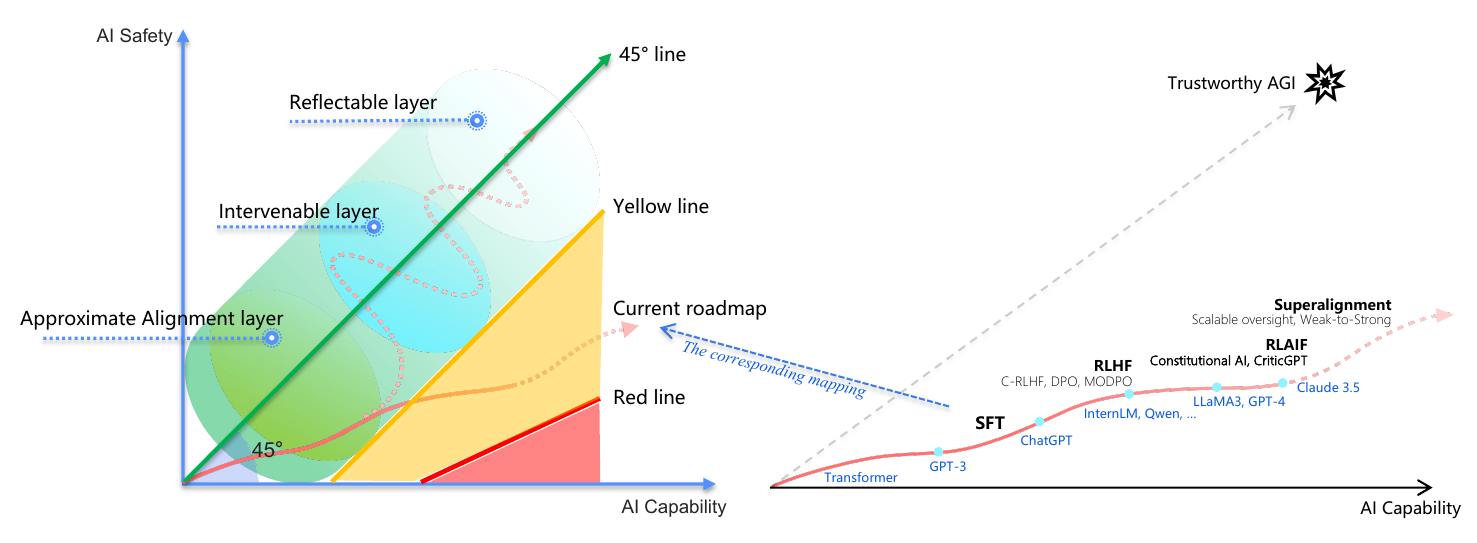}
      \caption{Left: Illustration of the AI-$45^{\circ}$ Law, which assumes that AI capability and safety should ideally be synchronized, represented by a $45^{\circ}$ line. Under the development of crippled AI, we can further divide the areas of existential risks (Red line) and early warning indicators (Yellow line). Right: Key milestone models in AI capability development.
      }
      \label{fig:45_degree}
\end{figure}

\subsection{AI-$45^{\circ}$ Law: Balancing Capability and Safety}
Recent advancements in AI have highlighted a significant gap between the rapid growth of AI capabilities and the slower progress in AI safety. 
In response to this imbalance, we introduce the AI-$45^{\circ}$ law as a guiding principle for the development of AI systems. This law posits that advancements in both AI capabilities and safety should ideally progress in parallel, with each dimension improving at the same rate—represented by a $45^{\circ}$ line in a capability-safety coordinate system, as shown in Figure \ref{fig:45_degree}. Although strict adherence to the $45^{\circ}$ line is not mandatory, some flexibility is allowed within a defined range. However, the current development trajectory deviates significantly from this ideal, with progress in AI safety lagging far behind the rapid acceleration of AI capabilities. 

\subsection{Red Line: Existential Risks and Catastrophic Consequences}
The unsafe development, deployment, or use of AI systems poses potential catastrophic risks \cite{bengio2024managing}, which could even become existential threats to humanity. As mentioned in the International Dialogues on AI Safety (IDAIS) \cite{idais-beijing}, they propose ``Red Lines'' of AI development in the following five aspects: \textit{autonomous replication or improvement; power-seeking; assisting weapon development; cyberattacks; deception}. At the same time, they also raise the need, including governance regimes and technical safety methods, to ensure these ``Red Lines'' will not be crossed. Following the AI-$45^{\circ}$ law, the ``Red Lines'' can be more clearly illustrated as occupying the lower right region relative to the $45^{\circ}$ line, as shown in Figure \ref{fig:45_degree}. These ``Red Lines'' - defined as those with the potential to lead to irreversible and catastrophic outcomes - are likely to increase as AI systems approach or surpass human-level intelligence.

\subsection{Yellow Line: Early Warning Indicators and Proactive Mitigation}
Considering the potential for extreme ``Red Line'' risks, it is critical to establish a framework for early-warning thresholds, called the ``Yellow Line'', as shown in Figure~\ref{fig:45_degree}, which can signal when the capabilities of an AI system approach a dangerous level. These thresholds would serve as indicators that a system may be on the verge of crossing into the ``Red Line'' territory. To achieve this, it is essential to build a scientific consensus on both the nature of AI risks and the appropriate boundaries that define these thresholds \cite{dalrymple2024towards}.

The concept of the ``Yellow Line'' is intended to complement and extend the existing safeguard assessment frameworks, such as responsible scaling policies \cite{rsp2023, rsp2024} from Anthropic. Models whose capabilities remain below these early-warning thresholds would require only basic testing and evaluation. However, for more advanced AI systems that exceed these thresholds, more rigorous assurance mechanisms and safety protocols would be necessary to mitigate potential risks \cite{idais-venice}. By establishing these thresholds, we can take a proactive approach to ensuring AI systems are developed, tested, and deployed with appropriate safeguards.
\section{The Causal Ladder of Trustworthy AGI}
\label{sec:ladder}
\begin{figure}[t]
\centering
\includegraphics[width=\textwidth]{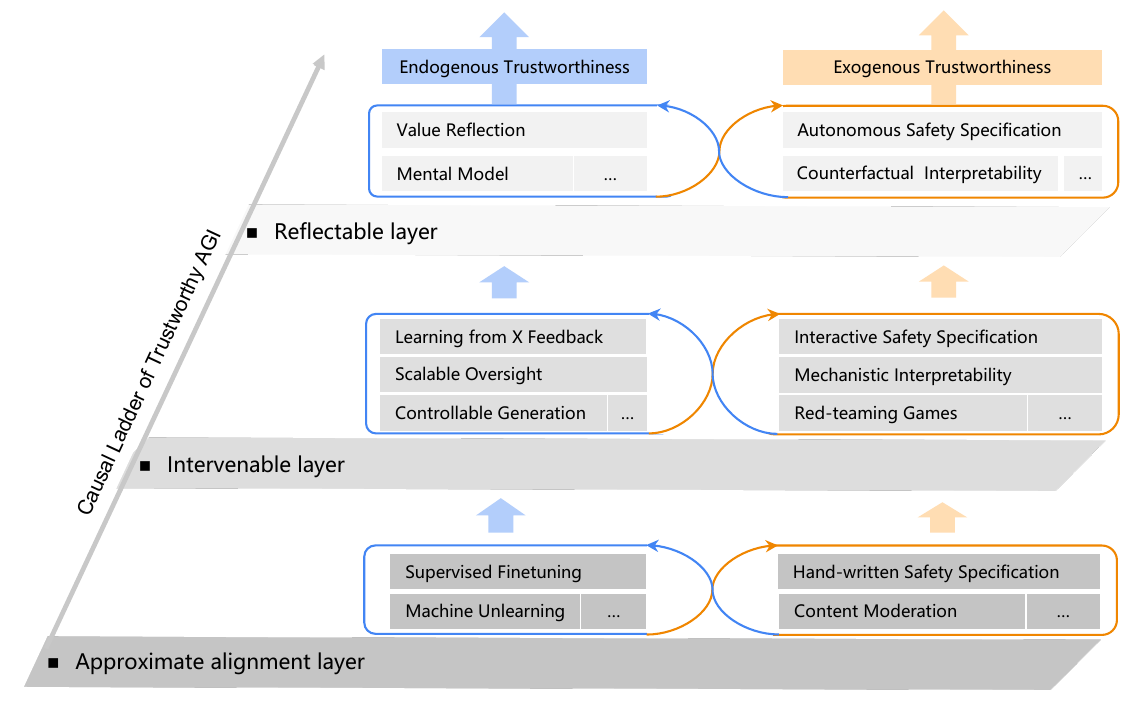}
\caption{
Illustration of the Causal Ladder of Trustworthy AGI: The framework consists of three core layers: \textit{Approximate Alignment}, \textit{Intervenable}, and \textit{Reflectable}. It integrates \textit{Endogenous Trustworthiness} and \textit{Exogenous Trustworthiness} to provide a comprehensive approach to ensuring AGI safety and trustworthiness. 
}
\label{fig:ladder}
\end{figure}

To explore a practical approach to the AI-$45^{\circ}$ law, we propose a three-layer technical framework based on the ``Ladder of Causation'' \cite{pearl2018book}, as illustrated in Figure \ref{fig:ladder}. This framework aims to address the critical safety and trustworthiness requirements on the path towards AGI \cite{li2023trustworthy, liu2023trustworthy, wang2023decodingtrust, huang2024trustllm}. Its core objective is to facilitate the progressive development of AGI systems, evolving from approximate alignment and intervention capabilities to self-reflection, thereby ensuring a high level of safety and trustworthiness. This approach not only overcomes the limitations of current AI models but also lays a technological foundation for future AGI development. 

Within this framework, we define two key dimensions of AGI trustworthiness: \textit{Endogenous Trustworthiness} and \textit{Exogenous Trustworthiness}, as shown in Figure \ref{fig:ladder}. Endogenous Trustworthiness refers to intrinsic safety technologies embedded within AGI systems, ensuring that safety mechanisms are inherent to the system's design and operation. In contrast, Exogenous Trustworthiness encompasses external mechanisms and assurance technologies that provide safety and reliability guarantees from an outside perspective.

\subsection{Layer 1: Approximate Alignment Layer} 

The first-layer technical approach focuses on the approximate alignment of AI models with human values. Currently, AI models exhibit limited generalization capabilities, where failing to accurately reflect human value systems, which may lead to biases and inconsistencies in real-world applications \cite{weidinger2023using}. Therefore, achieving broad alignment between models and human values is a crucial step in ensuring the safety and trustworthiness of AGI. This requires technical breakthroughs in areas such as value representation and alignment at the level of both conceptualization and encoding \cite{kong2024aligning}. By developing more nuanced and precise mechanisms for value representation, models can better comprehend and adhere to human values. Approximate alignment not only involves the model's ability to understand human instructions, but its capacity to make decisions that align with human ethics and morals in complex scenarios, thereby reducing the occurrence of biased behaviors. 
This serves as the foundational groundwork for the safe and trustworthy deployment of AGI, preventing actions that may conflict with human intentions during its operation.

This layer corresponds to the ``\textit{association}'' level of the ladder of causation. At this level, correlation-based techniques are applied to observational data, enabling AI models to answer the question ``\textit{What is it}''. In the context of developing trustworthy AGI, this corresponds to the \textit{Approximate Alignment Layer}, which involves data-driven approaches to extract and fit human values within a broad space. Key methods include, but are not limited to, supervised fine-tuning and machine unlearning.

\textbf{Supervised Fine-Tuning} \cite{dodge2020fine}: This method involves providing AI with high-quality, value-consistent question-and-answer data, facilitating the alignment of generative AI models, particularly those based on large language models, with human values through supervised learning techniques.

\textbf{Machine Unlearning} \cite{cooper2024machineunlearningdoesntthink, liu2024machine, dontsov2024clear, xing2024efuf, eldan2023s}: This technique aims to remove the influence of specific data from machine learning models, such as personal privacy data or erroneous information. By eliminating irrelevant patterns without compromising overall model performance, machine-unlearning effectively addresses issues related to data privacy and data leakage.

\subsection{Layer 2: Intervenable Layer}

The second-layer technological approach emphasizes addressing the need for safety verification and intervention during the model inference process. Most existing AI models, particularly deep learning-based black-box models, often lack transparency and interpretability in their inference mechanisms. This opacity makes external intervention and safety verification extremely challenging. To overcome this limitation, technological innovations must aim to decouple the inference process and fundamentally rethinking the reasoning mechanisms at the model architecture level. This approach ensures that each step of the inference process is traceable and verifiable.
Consequently, future AGI models should be designed to inherently interpretable and amenable to intervention, enabling human operators to monitor and modify the reasoning process in real time when necessary. These advancements would not only improve the safety of the model but also enhance understanding and optimization of their decision-making processes, providing a robust safety framework for deployment in complex environments.

This layer corresponds to the ``\textit{intervention}'' level of the ladder of causation, focusing on understanding and predicting the outcomes of interventions made on AI models. At this level, AI models address the question ``\textit{What will happen if X is intervened on?}''. Technologies relevant to this layer include intervention-based \cite{li2024inference} and reinforcement learning-based methods \cite{stiennon2020learning, ouyang2022training}, where interventions are simulated to study their effects or trial-and-error methods are used to influence the environment and optimize strategies through reinforcement learning \cite{sutton2018reinforcement}.

In the context of reliable AGI, the corresponding methodologies include, but are not limited to, feedback-based value alignment \cite{gabriel2020artificial} and scalable oversight \cite{bowman2022measuring, burns2023weak}, mechanistic interpretability \cite{bereska2024mechanistic}, and adversarial training \cite{perez2022red, hong2024curiosity}. We refer to this as the \textit{Interventable Layer}. At this level, external models or human participants provide feedback or interventions to guide and supervise the value alignment of large models during learning.

\textbf{Learning from X Feedback}: This involves providing correct value feedback from humans or AI-assisted humans based on the model's outputs and results, enabling human-in-the-loop reinforcement learning methods, such as reinforcement learning from human feedback (RLHF) \cite{rafailov2024direct, stiennon2020learning, ouyang2022training, zhou-etal-2024-beyond} and reinforcement learning from AI feedback (RLAIF) \cite{bai2022constitutional}.

\textbf{Controllable Generation} \cite{liang2024controllable}: Explicit control generation involves clearly defined instructions through human-computer interaction (e.g., input prompts), directing the model to generate text in a specific style, such as in a Shakespearean or humorous tone \cite{tao2024cat}. Implicit control generation \cite{zhou2024weak, liu2024inference}, on the other hand, refers to ensuring that the generated text meets certain standards even when
such requirements, such as producing non-toxic, inoffensive, and nondiscriminatory content, are not explicitly stated.

\textbf{Mechanistic Interpretability} \cite{bereska2024mechanistic,chen2024imitation,ren2024identifying,qian2024dean,zhang2024better,qian2024towards}: This approach involves intervening in the internal features or neuron weights of large models to observe the impact on model behavior or outcomes, thereby analyzing the model's safety performance, with a particular focus on feature factor analysis.

\subsection{Layer 3: Reflectable Layer}
The core of the third-layer technological pathway lies in overcoming the limitations of existing reasoning paradigms by introducing a novel reflective reasoning framework. 
Current models primarily depend on a ``chain of reasoning'', deriving conclusions based on existing inputs and experiences. However, these models often lack the capacity for genuine self-reflection \cite{renze2024self, shinn2024reflexion} and self-correction \cite{pan2023automatically}.
While this approach may be adequate for simpler tasks, it often falls to ensure safety and reliability in complex and dynamic real-world scenarios. Consequently, advancing AGI requires integrating self-reflective capabilities, enabling systems to evaluate their decisions throughout the reasoning process and adapt to environmental changes.
This ``reflective reasoning'' should operate not only within individual models but also extend to collaborative mechanisms among multiple models, allowing for mutual reflection and validation to enhance the credibility and safety of collective decision-making. By incorporating this reflective mechanism, systems can significantly improve robustness, mitigating critical errors arising from the accumulation of mistakes during continuous reasoning, and ultimately ensuring greater trustworthiness and reliability.

This layer corresponds to the ``\textit{counterfactual}'' level of the ladder of causation, enabling the AI to infer counterfactual scenarios by contemplating hypothetical conditions and evaluating outcomes under varying circumstances \cite{pearl2009causality,pearl2018book}. At this level, AI models address the question, ``What would have happened if a different choice had been made?'' In the development of trustworthy AGI, corresponding methods include, but are not limited to, world models \cite{ha2018recurrent}, value reflection \cite{pan2023automatically, shinn2024reflexion, renze2024self}, and counterfactual interpretability \cite{moraffah2020causal, verma2020counterfactual,van2021interpretable,chen2024quantifying,chen2024cello}, collectively referred to as the reflectable layer.

\textbf{Value Reflection} \cite{pan2023automatically, shinn2024reflexion, renze2024self}: This process involves AI engaging in a deep deliberation about its actions and choices. Through this reflective process, AI systems can determine which values are important and worth pursuing. By continuously reflecting on and optimizing its understanding of human values and preferences, the AI system can better align with and embody these values, improving its safety, reliability, and societal acceptance.

\textbf{Mental Models} \cite{forrester1971counterintuitive, gentner2014mental,chen2024imitation}, or called world models \cite{ha2018recurrent}: 
World models offer a mathematical representation of how an AI system interacts with and influences the external environment. These models enable AI systems to predict the downstream effects of their decisions and comprehend the broader implications of their actions, fostering more informed and responsible decision-making. Unlike data correlation analysis, world models empower AI systems to engage in counterfactual reasoning and explore "\textit{what if}" scenarios—capabilities that are natural for humans but challenging for current AI technologies. Advancements in this area would substantially enhance AI's decision-making capabilities.

\textbf{Counterfactual Interpretability} \cite{moraffah2020causal, verma2020counterfactual,van2021interpretable,chen2024quantifying,chen2024cello}: This aspect focuses on causal attribution, mediation analysis, and related techniques, aiming to provide a clear understanding of the underlying causal mechanisms and their implications for the decision-making processes.

\subsection{Implications}

These technological innovations will not only provide critical support for AGI development but will also address its potential social impacts and security risks. In future applications, AGI will progressively achieve safety and trustworthiness, ensuring balanced development of capabilities under the premise of safety and trust. This series of breakthroughs will guide AGI toward a more mature stage, becoming a key driver of societal progress while safeguarding against uncontrolled risks to humanity during its development.

\section{The Matrix of Trustworthy AGI}
\begin{figure}
\centering
\includegraphics[width=\textwidth]{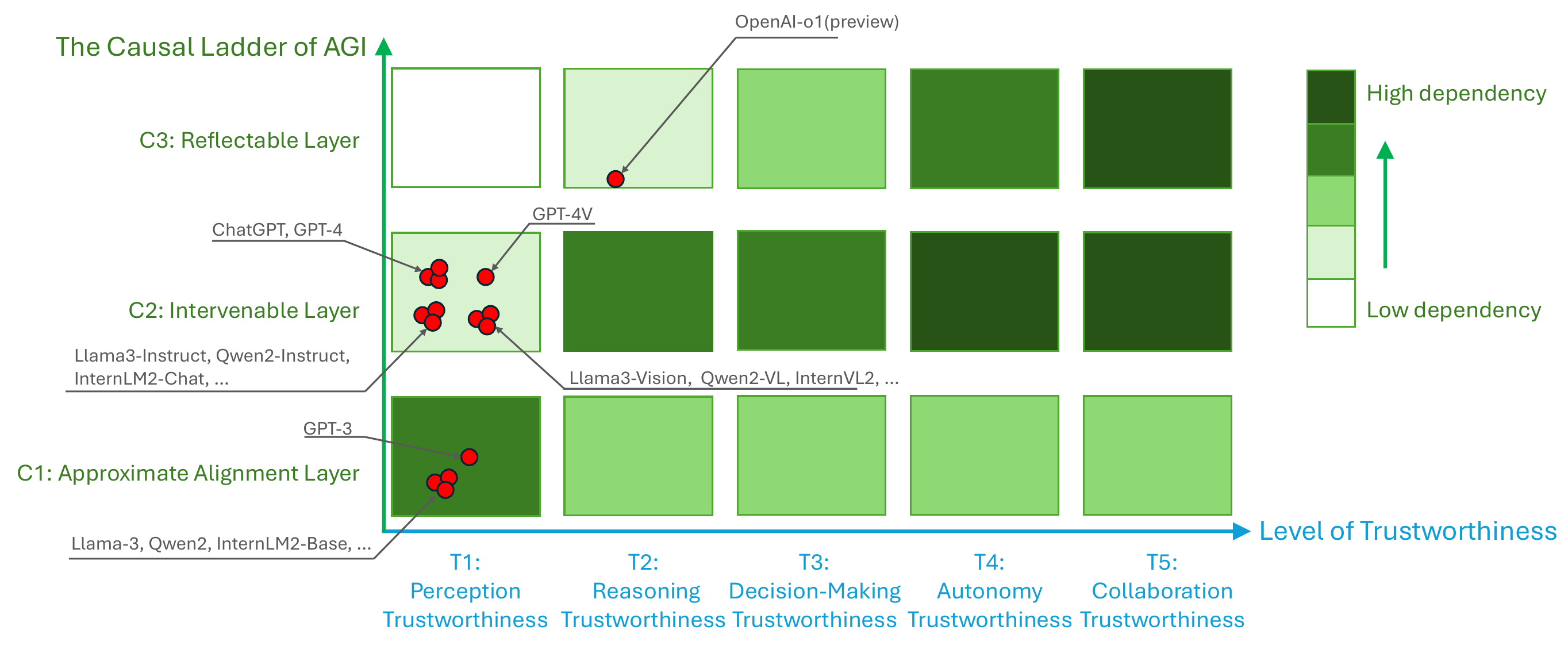}
\caption{Matrix of trustworthy AGI: Based on the causal ladder of AGI and the levels of trustworthiness, we illustrate the positions of several representative models within the matrix. Reliance on the Reflectable Layer increases as we progress through the levels of trustworthiness.
}
\label{fig:matrix}
\end{figure}

Based on the causal ladder of trustworthy AGI, we prospectively define five levels of AGI trustworthiness: \textit{Perception Trustworthiness}, \textit{Reasoning Trustworthiness}, \textit{Decision-making Trustworthiness}, \textit{Autonomy Trustworthiness}, and \textit{Collaboration Trustworthiness}. 
These levels collectively form a comprehensive framework for a trustworthy AI system, supporting the balanced development of trustworthiness and capabilities in AGI.
The rationale behind this taxonomy is to systematically address the different facets of trustworthiness in AGI systems, ensuring that each foundational layer contributes to the overall reliability and ethical alignment of the AI. Each level builds upon the previous one, creating a hierarchical structure that enhances the AGI's ability to operate effectively and ethically in complex environments.

As illustrated in Figure \ref{fig:matrix}, the dependence on each layer of the causal ladder for ensuring trustworthy AGI varies across levels. As we progress through the levels of AGI trustworthiness, reliance on the Reflectable Layer increases. For example, achieving Perception Trustworthiness does not necessitate the techniques from the Reflectable Layer, whereas Reasoning Trustworthiness does, albeit to a limited extent.

As reported by current large language model techniques, a series of foundation models—such as Llama2/3 \cite{touvron2023llama, dubey2024llama}, Qwen \cite{bai2023qwen, yang2024qwen2}, and InternLM \cite{cai2024internlm2} -- utilize autoregressive methods to pre-train on vast amounts of safety-filtered text data. These models are then fine-tuned on specific tasks through supervised learning, representing techniques within the Approximate Alignment Layer. This approach has yielded impressive results in text generation.
Furthermore, through human-in-the-loop methods, such as reinforcement learning \cite{rafailov2024direct, stiennon2020learning, ouyang2022training} can be applied across various tasks to align with human values, placing these methods within the Intervenable Layer and achieving even better performance.
These techniques can also be extended to multimodal tasks, such as image understanding \cite{bai2023qwenvl, wang2024qwen2vl, chen2024internvl, dong2024internlm-xcom2}, image generation \cite{wang2024emu3}, and video generation \cite{openai-sora}. However, at their core, they still focused on the Perception Trustworthiness level.

As one of the most advanced reasoning models, OpenAI-o1 (preview) \cite{openai-o1} is classified as a more basic form of the Reflectable Layer. Due to its incorporation of preliminary policy review and safety verification in the reasoning process, we also consider it to fall within the level of Reasoning Trustworthiness.

\paragraph{Level 1: Perception Trustworthiness.}

Perception trustworthiness refers to the reliability and accuracy of the AI system in gathering, processing, and interpreting sensory data from its environment. This level ensures that AI can make consistent and accurate observations about the world, free from perceptual biases or inaccuracies. By providing trustworthy inputs, it enables downstream reasoning and decision-making processes to be based on valid and reliable data.

\paragraph{Level 2: Reasoning Trustworthiness.}
Reasoning trustworthiness involves the AI system's ability to perform logical, causal, or probabilistic reasoning in a manner that is transparent, consistent, and verifiable. This level guarantees that the AI's reasoning steps are understandable, traceable, and aligned with predefined ethical standards or domain-specific principles. It includes mechanisms for verifying intermediate results and ensuring robustness against errors during inference, thereby enhancing confidence in the AI's cognitive processes.

\paragraph{Level 3: Decision-making Trustworthiness.}
Decision-making trustworthiness pertains to the transparency, consistency, and value alignment of the AI's decision-making process, particularly within embodied AI systems that interact with the physical world. At this level, decision-making must be timely and context-aware, ensuring that actions in response to environmental stimuli align with human values and ethical standards. Trustworthiness here guarantees that the AI's decisions adhere to clearly defined ethical frameworks and are accompanied by explanations that render the decision process understandable to humans. Additionally, it incorporates mechanisms for monitoring, intervention, and adjustment to ensure that decisions are goal-directed, safe, and ethically aligned, thereby enabling effective operation in real-world, dynamic settings while earning human trust.

\paragraph{Level 4: Autonomy Trustworthiness.}
Autonomy trustworthiness refers to the AI's ability to self-regulate during autonomous operations while maintaining alignment with ethical principles and specified goals. This level includes mechanisms for reflection, self-improvement, and self-constraint to ensure that the AI's autonomous actions do not deviate from ethical boundaries. It ensures that the AI can independently adapt and plan actions in dynamic environments without compromising its core values and alignment, thereby sustaining trust during independent operations.

\paragraph{Level 5: Collaboration Trustworthiness.}
Collaboration trustworthiness focuses on the AI’s capacity to work effectively and transparently in multi-agent environments, including interactions with both humans and other AI systems. It encompasses the establishment of clear interaction rules, threat models to prevent conflicts of interest, and mechanisms for reaching consensus in value negotiations. This level ensures that the AI collaborates in a stable, ethical, and goal-aligned manner, maintaining reliability even in complex, highly dynamic environments.

Overall, achieving trustworthy AGI necessitates a comprehensive approach that addresses all five levels. Perception Trustworthiness ensures accurate and reliable sensory data collection, forming the foundation for all subsequent processes. Reasoning Trustworthiness demands transparent and explainable reasoning processes, maintaining logical consistency and causal reasoning abilities to provide reliable results in complex tasks. Decision-making Trustworthiness requires that the AI's decisions align with human values, effectively handle uncertainty, and exhibit dynamic adaptability to ensure rationality and stability. Autonomy Trustworthiness involves the AGI's capacity for self-reflection and self-correction, continuously optimizing its decision-making processes through ongoing learning to ensure safety and reliability in independent tasks. Finally, Collaboration Trustworthiness emphasizes transparency and reliability in both human-machine and multi-agent collaborations, ensuring effective information sharing and cross-verification between systems to minimize risks.

By systematically developing and integrating trustworthiness at each of these levels, AGI systems can achieve a balanced integration of capability and trustworthiness, ultimately fostering human trust and facilitating beneficial human-AI collaboration.
\section{Governance measures}
\label{sec:openproblems}

\textbf{Lifecycle management} \cite{judge2024code, novelli2024accountability, hogberg2024stabilizing,huang2024pixels}: Ensuring effective AI governance throughout the entire lifecycle—from development to deployment and eventual decommissioning—poses challenges in maintaining accountability, transparency, and adaptability in rapidly evolving technologies.

\textbf{Multi-stakeholders} \cite{de2021artificial, van2013value}: AI governance must navigate the complexities of balancing diverse stakeholder interests, including governments, corporations, academia, and civil society, while ensuring fair representation, collaboration, and accountability in decision-making processes.

\textbf{Governance for good} \cite{palladino2023biased, huang2024pixels}: Achieving ethical AI governance that promotes societal well-being requires overcoming challenges related to aligning AI systems with human rights, mitigating biases, and preventing misuse, while also fostering innovation and public trust.

\textbf{AI safety as a global public good} \cite{hendrycks2023overview, globalgoods}: AI safety is increasingly recognized as a global public good due to the rapid advancement of AI systems that may soon surpass human intelligence. While these systems promise great potential, they also pose catastrophic risks if misused or uncontrollable. Given the global nature of these threats, it is crucial to establish effective governance and safeguard mechanisms to mitigate these risks and ensure humanity's security.

\section{Conclusion}
In this paper, we have introduced the AI-\textbf{$45^{\circ}$} Law aimed at balancing the development of AI safety and capability. Central to our contribution is the Causal Ladder of Trustworthy AGI, a practical framework that offers a technical taxonomy for existing research methodologies. Our framework consists of three core layers: the Approximate Alignment Layer, the Intervenable Layer, and the Reflectable Layer. Each layer addresses specific aspects of safety and trustworthiness in AGI systems.

We have also defined five progressive levels of AGI trustworthiness: Perception Trustworthiness, Reasoning Trustworthiness, Decision-making Trustworthiness, Autonomy Trustworthiness, and Collaboration Trustworthiness. These levels collectively form a comprehensive framework for developing AGI systems that are not only highly capable but also aligned with human values and ethical standards. By systematically categorizing and addressing the challenges at each level, our framework facilitates a balanced approach to advancing both the capability and the safety of AGI.

Despite these advancements, several challenges and open problems persist. Refining the methodologies within each layer of the causal ladder, integrating ethical considerations more deeply into the framework, and ensuring adaptability in dynamic real-world environments are areas that require further research. Moreover, fostering collaboration among researchers, policymakers, and industry stakeholders is essential for addressing the multifaceted issues surrounding trustworthy AGI.

Future work will empirically validate the proposed framework and explore its applicability across domains and AI architectures. By continuing to develop and refine this framework, we aim to contribute to the creation of AGI systems that are not only powerful and efficient but also safe and trustworthy. This balanced approach is imperative for harnessing the full potential of AGI while mitigating risks, ultimately leading to technological advancements that benefit society as a whole.

\section*{Acknowledgements}
We sincerely thank each member of the Center for Safe \& Trustworthy AI at the Shanghai Artificial Intelligence Laboratory for their invaluable contributions and suggestions, which made this position paper possible. We also deeply appreciate the input from the broader AI safety and trustworthiness community.

\bibliography{
references/1-progress,
references/2-risks,
references/3-crippled_ai,
references/4-law,
references/5-ladder
}

\clearpage
\appendix

\clearpage

\end{document}